\documentstyle[seceq,mbf,epsf]{ptptex}
\def\dA{d_{\rm A}}

\def\half{{1\over 2}}
\newcommand{\dAi}{d_{{\rm Ai}}}

\newcommand{\dAf}{d_{{\rm A}1}}
\newcommand{\la}{\lambda}
\newcommand{\sAB}{{\sigma_{AB}}}

\newcommand{\beq}{\begin{equation}}
\newcommand{\eeq}{\end{equation}}
\newcommand{\bea}{\begin{eqnarray}}
\newcommand{\eea}{\end{eqnarray}}
\newcommand{\br}{{\bf r}}
\newcommand{\be}{{\bf e}}
\newcommand{\bn}{{\bf n}}
\newcommand{\bl}{{\bf l}}
\newcommand{\cd}{{\cal D}}
\begin{document}
 \makeatletter
 \newdimen\ex@
 \ex@.2326ex
 \def\dddot#1{{\mathop{#1}\limits^{\vbox to-1.4\ex@{\kern-\tw@\ex@
  \hbox{\tenrm...}\vss}}}}
 \makeatother
\notypesetlogo  
\markboth{
N.~Sugiura et al.
}{
How Do Nonlinear Voids Affect Light Propagation ?
}
\title{
How Do Nonlinear Voids Affect Light Propagation ?
}

\author{
Norimasa {\sc SUGIURA}$^{1,}$\footnote{E-mail:
  sugiura@tap.scphys.kyoto-u.ac.jp} 
Ken-ichi {\sc NAKAO}$^{2,}$\footnote{E-mail: 
  knakao@sci.osaka-cu.ac.jp}
Daisuke {\sc IDA}$^{1,}$\footnote{E-mail:
  ida@tap.scphys.kyoto-u.ac.jp}
Nobuyuki {\sc SAKAI}$^{3,}$\footnote{E-mail:
  sakai@yukawa.kyoto-u.ac.jp}
and
Hideki {\sc ISHIHARA}$^{4,}$\footnote{E-mail:
  ishihara@th.phys.titech.ac.jp}
}

\inst{
$^{1}$Department of Physics, Kyoto University, Kyoto 606-8502, Japan
\\
$^{2}$Department of Physics, Osaka City University, Osaka 558-8585, Japan
\\
$^3$Yukawa Institute for Theoretical Physics, Kyoto University\\
Kyoto 606-8502, Japan
\\
$^4$Department of Physics, Tokyo Institute of Technology\\
Tokyo 152-0033, Japan
}
\recdate{September 29, 1999
}
\abst{
Propagation of light in a clumpy universe is examined.
As an inhomogeneous matter distribution,
we take a spherical void surrounded by a dust shell,
where the ``lost mass'' in the void is compensated by the shell.
We study how the angular-diameter distance behaves
when such a structure exists.
The angular-diameter distance is calculated
by integrating the Raychaudhuri equation including the shear.
An explicit expression for the junction condition for the 
massive thin shell is calculated. We apply these results 
to a dust shell embedded in a Friedmann universe
and determine how the 
distance-redshift relation is modified compared with that in the purely 
Friedmann universe.
We also study the distribution of distances in a universe filled with voids.
We show that the void-filled universe
gives a larger distance than the FRW universe
by $\sim 5\%$ at $z \sim 1$ if the size of the void 
is $\sim 5\%$ of the Horizon radius.
}
\maketitle
\makeatletter
\if 0\@prtstyle
\def\asp{.3em} \def\bsp{.26em}
\else
\def\asp{.3em} \def\bsp{.3em}
\fi \makeatother
\section{Introduction}
One of the most interesting findings of the nearby redshift surveys
is the discovery of large voids and that they appear
to be a common feature of galaxy distributions.
Great interest in such voids was first aroused by the discovery 
of the Bo\"{o}tes void,\cite{rf:Kirshner1}\ 
which was confirmed to have a radius of $\sim 60~h^{-1}{\rm Mpc}$. 
\cite{rf:Kirshner2}\ 
At that stage, however, 
it could not be concluded that voids are common structures
in the universe. 
The idea that voids are in fact common features of the large-scale 
structure of our universe was established by the recent surveys 
beginning from the CfA redshift survey,\cite{rf:deLapparent}\ 
which introduced a picture of the universe where 
the galaxies are located on the surfaces of 
bubble-like structures with diameters $25-50~{\rm Mpc}$. 
Using the survey data such 
as CfA2\cite{rf:GELLER} and SSRS2,\cite{rf:DACOSTA}\ 
El-Ad {\it et al.},\cite{rf:ELAD}\  
who established a method to find voids in these galaxy distributions,
revealed that a substantial fraction of the volume of our universe is
occupied by underdensity regions; 
they have shown that about half of
the volume is filled with voids and that the voids have a diameter of
at least 40$h^{-1}$ Mpc with an average underdensity of $- 0.9$.

With the motivation of explaining such large voids, the gravitational
evolution of a less-dense region has been studied by several authors.
\cite{rf:PEEBLES,rf:SATO1}\ It was shown \cite{rf:SATO1} that
a less-dense region expands faster than the outer region, and a dense thin
shell is formed behind the shock front by the ``snow-plow'' mechanism.
The general relativistic motion of the void's shell was also studied by Maeda
and Sato.\cite{rf:MAEDA}\ They adopted the
metric junction method developed by Israel \cite{rf:ISRAEL}\ and derived
an expansion law of a void in the Friedmann-Robertson-Walker (FRW) 
universe. 
In addition to these, non-gravitational scenarios for the formation of voids
have also been proposed. For example, in the explosion model,
\cite{rf:OSTRIKER}\ a void surrounded by a thin shell is formed
by the explosion of a pregalactic object.
Another example is given by an inflationary model with a
first-order 
phase transition, where vacuum bubbles are nucleated during inflation.
These bubbles could represent the origin of voids.\cite{rf:AMENDOLA}\
Thus the void model not only provides a simple picture of nonlinear density
fluctuations but also is supported by many observational and theoretical
considerations.

There are also studies which investigate light propagation properties in a
void system, such as
the observational effect of voids on CMB anisotropies
and the modification of the redshift and luminosity for distant sources
due to the existence of voids.
The effects on the CMB anisotropies
were discussed by several authors,\cite{rf:THOMPSON,rf:SAKAI}\ 
suggesting that further CMB observations will give some
constraints on the configuration and origin of voids. 
Investigation of 
the light propagation properties in a clumpy universe
using a spherical void model 
was undertaken by Sato,\cite{rf:SATO} \ 
who discussed the modification of the redshift and luminosity in
comparison with the FRW model. He found that the modification is 
third order in ($H r_v/c $) for the redshift and 
first order in ($H r_v/c $) for
the luminosity, where $H$ is the Hubble parameter, $r_v$ is the
size of the void and $c$ is the velocity of light.
We believe, however, his result is suspect
because the condition for the distance junction he adopted 
seems to be inappropriate.

In this paper, we study light propagation in a void system
based on the Raychaudhuri equation. A light ray bundle traveling
across a spherical void surrounded by a thin shell suffers
(de-)magnification and shear focusing. We calculate the modification
of the distance and redshift in comparison with the FRW model.
Light propagation in the Swiss Cheese model has been 
studied extensively,\cite{rf:DYER,rf:WALD}\ 
but studies on a void surrounded by an overdense shell
are far from complete.
This paper gives an exact treatment of such a
system. 

The organization of this paper is as follows.
In {\S}2 we derive the junction condition for the expansion and
shear of a null ray bundle across a shell.
In {\S}3 we calculate the modification of the 
distance and the redshift
caused by the void-shell system and 
discuss its cosmological implications.
The last section is devoted to a summary.
We follow the signature of the metric and the convention of the
Riemann tensor used in Ref. 18).

\section{Basic equations}
In this paper, we consider a spherical void in the FRW universe. 
We assume that the interior of the void is a vacuum,
so that the spacetime inside the void is flat.
The mass that would have been in the void 
is assumed to be distributed in a spherical shell
surrounding the void.
We treat the matter in the shell as an infinitely thin dust shell.

In this section, we derive an expression for the angular-diameter distance 
in a flat spacetime and in FRW spacetime, which 
respectively correspond to the inside and outside spacetime of
the void. 
We also derive the junction condition for the expansion and shear 
of a light ray bundle across the dust shell.
\subsection{Basic relations in geometric optics}
We use the notation used by
Sasaki.\cite{rf:SASAKI}\ 
Let $k^{a}$ be the tangent of a null geodesic
parameterized by an affine parameter, $\lambda$.
We introduce a null vector field, $l^a$, defined by
\begin{eqnarray}
  l^a l_a =0,~~l^a k_a =-1,~~{l^a}_{;b} k^b =0 ,
\end{eqnarray}
where the semi-colon denotes the covariant derivative.
Using these vectors, we define the metric $H_{ab}$ on the two-dimensional
screen placed orthogonal to the spatial direction of the null
geodesics:
\begin{eqnarray}
  H_{ab} \equiv g_{ab} + k_a l_b + l_a k_b  .
\end{eqnarray}
The expansion $\theta$ and the shear $\sigma$ of the cross section of
the beam are defined by 
\begin{equation}
\theta = {k^{a}}_{;a},~~~
\sigma_{ab} = {H_{a}}^{c} \left( k_{c;d}-k_{d;c}
-\half H_{cd}\right){H_b}^{d}.
\end{equation}
We also introduce a dyad, $\{ e_{A}^{a} \}$ ($A=1,~2$), 
on the screen such that
\begin{eqnarray}
  H_{ab} = \delta_{AB} {e_{a}}^{A}{e_{b}}^{B},~~
  g^{ab}{e_{a}}^{A}{e_{b}}^{B} = H^{ab}{e_{a}}^{A}{e_{b}}^{B}=
 \delta^{AB}~~{\rm and}~~
e^{a}_{A;b} k^b = 0 ;
\end{eqnarray}
i.e. they are parallel-propagated along the geodesics.
Then the basic equations of the geometric optics can be written as
\cite{rf:SACHS}
\begin{eqnarray}
{d\theta\over d\lambda}&=&-R_{44}-\sigma^{AB}\sigma_{AB}
-{1\over2}\theta^{2}, \label{eq:expansion-eq}\\
{d\sigma_{AB}\over d\lambda}
&=&-C_{A4B4}-\theta\sigma_{AB},\label{eq:shear-eq}
\end{eqnarray}
with
\begin{equation}
\sigma_{AB} = \sigma_{ab} e^{a}_{A} e^{b}_{B},~~~
R_{44}\equiv R_{ab}k^{a}k^{b},~~~
C_{A4B4}\equiv C_{abcd}e_{A}^{a}k^{b}e_{B}^{c}k^{d},
\end{equation}
where $R_{ab}$ and $C_{abcd}$ are the Ricci and Weyl tensor,
respectively. 
Finally, we define the angular-diameter distance $\dA$ by
\begin{eqnarray}\label{def:dA}
  {d\over d\la}\ln{\dA} = \half \theta .
\end{eqnarray}
By integrating Eqs. (\ref{eq:expansion-eq}), (\ref{eq:shear-eq}) and
(\ref{def:dA}), we can obtain the angular-diameter distance.
We can also write the equation for $\dA$ as
\begin{eqnarray}
  {d^2\over d{\la}^2}\dA {\Big /}\dA = -\half R_{44}
    -{1\over 2}\sigma^{AB}\sigma_{AB}.
\end{eqnarray}
This expression is especially useful when the shear term vanishes.

\subsection{Inside the void---flat spacetime}
In flat spacetime,
the Raychaudhuri equations reduce to
\begin{eqnarray}\label{Ray-Min}
  {d\over d\lambda}\theta = - \half \theta^2 - \sigma^{AB}\sigma_{AB}
\end{eqnarray}
and
\begin{eqnarray}\label{Ray2-Min}
  {d\over d\lambda}\sAB = -\theta \sAB .
\end{eqnarray}
Suppose that the expansion $\theta$ is positive initially,
which implies that the derivative of $\dA$ is positive.
Then the angular diameter-distance is given by
\begin{eqnarray}\label{dA-Min}
  \dA &=&  {\dA}_0 \left\{1+\half \theta_0 (\la - \la_0)\right\}
\sqrt{ 1 - 
\left\{ 
  {|\sigma_{AB0}| (\la - \la_0)\over 1+\half \theta_0 (\la - \la_0)}
\right\}^2},
\end{eqnarray}
where $|\sigma_{AB0}| \equiv |\sigma_{AB0}\sigma^{AB}_0|^{1/2}$.
Note that when the shear vanishes, the distance is proportional to the
time coordinate:
\begin{eqnarray}
  {d \over dt}\dA = ~{\rm constant} .
\end{eqnarray}
By inspecting Eq. (\ref{dA-Min}), one can see that the shear term 
appears as a correction to the shear-free case.

\subsection{Jumps of the optical scalars across the 
singular hypersurface}
The shell which surrounds the void generates a timelike hypersurface
through its motion.
Since we treat the shell as infinitely thin, the hypersurface
may be regarded as singular.
We follow the procedure to treat a singular
hypersurface formulated by Israel.\cite{rf:ISRAEL}\ 
In order to obtain the jumps of the expansion and the shear
of a null geodesic bundle across the shell, we proceed as follows.

We consider a spacetime $V$ with a timelike singular hypersurface 
$\Sigma$.
The singular hypersurface $\Sigma$ divides $V$ into two regions, 
$V_{+}$ and $V_{-}$. 
Consider a field variable $\Psi$ defined on $\Sigma$.
The values of $\Psi$ evaluated on either side of $\Sigma$ 
will be denoted by $\Psi_{\pm}$ and  
the jump is defined by 
\begin{equation}
\left[\Psi\right]\equiv \Psi_{+} - \Psi_{-}.
\end{equation}
We introduce a spacelike unit 
vector field $n^{a}$ ($n^{a}n_{a}=+1$) normal 
to $\Sigma$ and the projection operator 
$h^{a}_{b}\equiv \delta^{a}_{b}-n^{a}n_{b}$. 
The extrinsic curvature $K_{ab}$ is then given by
\begin{equation}
K_{ab}=-{1\over2}h_{a}^{c}h_{b}^{d}\pounds_{n}h_{cd}
=-h_{a}^{c}h_{b}^{d}\nabla_{c}n_{d},
\end{equation}
where $\pounds_{n}$ and $\nabla_{a}$ are the Lie derivative along
$n^{a}$ and the covariant derivative, respectively. 
We have two extrinsic curvatures, $K_{ab}|_{\pm}$, 
which may differ, since $\Sigma$ is singular. 
The jump of $K_{ab}$ across $\Sigma$ can be written in the form
\begin{equation}
\left[K_{ab}\right]=8\pi G
\left(S_{ab}-{1\over2}h_{ab}S^{c}_{c}\right). \label{eq:j-condition}
\end{equation}
The tensor $S_{ab}$ here is related to the stress-energy tensor $T_{ab}$ 
through the equation
\begin{equation}
S_{ab}=\lim_{\varepsilon\rightarrow 0}\int_{-\varepsilon}^{+\varepsilon}
T_{cd}h^{c}_{a}h^{d}_{b}dx,
\end{equation}
where $x$ is a Gaussian normal coordinate ($x=0$ on $\Sigma$). 
Hence $S_{ab}$ is regarded as the surface stress-energy tensor of 
$\Sigma$. 

The jumps of $\theta$ and $\sigma_{AB}$ on $\Sigma$ are
calculated by integrating Eqs.(\ref{eq:expansion-eq}) and (\ref{eq:shear-eq})
 as 
\begin{eqnarray}
 \left[\theta\right]
&=&-\lim_{\epsilon\rightarrow 0_{+}}
 \int_{-\epsilon}^{+\epsilon}R_{44}\ dx / k^{\rm n},
\label{eq:theta-jump} \\
 \left[\sigma_{AB}\right]
&=&-\lim_{\epsilon\rightarrow 0_{+}}
 \int_{-\epsilon}^{+\epsilon} C_{A4B4}\ dx / k^{\rm n},
\label{eq:sigma-jump}
\end{eqnarray}
where 
\begin{eqnarray}
k^{\rm n} \equiv k^{a}n_{a}. 
\end{eqnarray}
Here we have set $n^{a}$ to point in the direction in which 
the affine parameter increases. From 
the above equations, we see that the jumps of $\theta$ 
and $\sigma_{AB}$ on $\Sigma$ come from the diverging parts of 
$R_{44}$ and $C_{A4B4}$ on $\Sigma$. 
Thus our next task is to calculate $R_{44}$ and $C_{A4B4}$.
After a straightforward manipulation, we find 
\begin{eqnarray}
 R_{44}
&=&(k^{\rm n})^{2}\pounds_{n}K^{c}_{c}
 +k^{a}k^{b}h_{a}^{c}h_{b}^{d}\pounds_{n}K_{cd}+
 {\rm finite}~~{\rm term}, \label{eq:R44}\\
 C_{A4B4}
&=&\left\{ \left(e_{A}^{\rm n}e_{B}^{\rm n}
 -{1\over2}\delta_{AB}\right)k^{a}k^{b}
 -k^{\rm n}\left(e_{A}^{\rm n}e_{B}^{a}
 +e_{B}^{\rm n}e_{A}^{a}\right)k^{b}
 +(k^{\rm n})^{2}e_{A}^{a}e_{B}^{b}
 \right\} \nonumber \\
&\times& h_{a}^{c}h_{b}^{d}\pounds_{n}K_{cd}
 -{1\over2}\delta_{AB}(k^{\rm n})^{2}\pounds_{n}K^{c}_{c}
 +{\rm finite}~~{\rm term} ,\label{eq:CA4B4}
\end{eqnarray}
where $e_{A}^{\rm n} \equiv e_{A}^a n_a$.
Using Eqs. (\ref{eq:j-condition}) and
(\ref{eq:theta-jump})--(\ref{eq:CA4B4}), 
we finally obtain the jumps of $\theta$ and $\sigma_{AB}$ as
\begin{eqnarray}
\left[\theta\right]
&=&-{8\pi G\over k^{\rm n}}k^{a}k^{b}S_{ab}, \label{eq:theta-jump2}\\
\left[\sigma_{AB}\right]
&=&-{8\pi G\over k^{\rm n}}
 \left\{\left(e_{A}^{\rm n}e_{B}^{\rm n}
 -{1\over2}\delta_{AB}\right)k^{a}k^{b}
 -k^{\rm n}\left(e_{A}^{\rm n}e_{B}^{a}
 +e_{B}^{\rm n}e_{A}^{a}\right)k^{b} 
 +\left(k^{\rm n}\right)^{2}e_{A}^{a}e_{B}^{b}\right\} \nonumber \\
&\times& \left(S_{ab}-{1\over2}h_{ab}S^{c}_{c}\right)
 -2\pi G\delta_{AB}k^{\rm n}S_{c}^{c}. \label{eq:sigma-jump2}
\end{eqnarray}

In this paper, 
we assume that the singular hypersurface which approximates the 
wall of the void has a surface stress-energy tensor of the dust type
(we call it a dust shell),
\begin{equation}
S_{ab}=Du_{a}u_{b},
\end{equation}
where $D$ is the surface density and $u^{a}$ is the 4-velocity 
of the shell ($u^{a}u_{a}=-1$). 
In the case of the dust shell, 
Eqs. (\ref{eq:theta-jump2}) and 
(\ref{eq:sigma-jump2}) become 
\begin{eqnarray}
 \left[\theta\right]
&=&-{8\pi G \over k^{\rm n}}D\left(k^{\rm u}\right)^{2}, 
\label{eq:theta-jump3}\\
 \left[\sigma_{AB}\right]
&=&-{4\pi G \over k^{\rm n}}D
 \Bigl[2\bigl(e_{A}^{\rm n}k^{\rm u}
 -e_{A}^{\rm u}k^{\rm n}\bigr)
 \bigl(e_{B}^{\rm n}k^{\rm u}
 -e_{B}^{\rm u}k^{\rm n}\bigr) \nonumber \\
&&~~~~~~~~~~~~~~~~~ +\delta_{AB}
 \bigl\{(k^{\rm n})^{2}-(k^{\rm u})^{2}\bigr\}
\Bigr],\label{eq:sigma-jump3}
\end{eqnarray}
where
\begin{equation}
k^{\rm u}\equiv k^{a}u_{a}~~~~{\rm and}~~~~
e_{A}^{\rm u}\equiv e_{A}^{a}u_{a}.
\end{equation}

\subsection{Outside the void---FRW spacetime}
We note here the analytical expression for the
distance-redshift relation in flat FRW spacetime.
The basic assumption is that the shear is negligible.
We find the expression for $\dA$ in terms of the cosmic time $t$ as 
\begin{eqnarray}
 \dA = \dAi \left\{ 3 \left( {t \over t_{\rm i}} \right)^{2/3}-
   2\left( {t \over t_{\rm i}} \right)\right\}
+ {t_{\rm i}\over 2k_{\rm i}} {\dAi} \theta_{\rm i}
\left\{ 3 \left( {t \over t_{\rm i} }\right)-
    3\left({t \over t_{\rm i}} \right)^{2/3}\right\}, \label{dA-FRW}
\end{eqnarray}
with an appropriate boundary condition for $\theta_{\rm i}$,
where $k_{\rm i}$ is the frequency of the ray 
observed by a comoving observer at $t= t_{\rm i}$.
\section{Application and discussion}
\subsection{Single void}
We consider a spherical void surrounded by a dust shell in a 
flat FRW universe (which we refer to in the following as a ``FRW
model''). The matter that would have been inside the void is
distributed in the shell and we assume that the external universe is
undisturbed. This is exactly the same as the model analyzed by Thompson and
Vishniac.\cite{rf:THOMPSON}\ We consider a light ray bundle traveling
across the void and calculate the redshift and the angular-diameter
distance using the junction conditions obtained in the previous
section. The result will be compared with the FRW relation.

The metric of the FRW model
is given by
\begin{eqnarray}
  ds^2 = - dt^2 + a(t)^2(dr^2+r^2 d\Omega^2).
\end{eqnarray}
The void is an empty region which can be described by a
flat metric,
\begin{eqnarray}
  ds^2 = - dt'^2 + dx'^2 +x'^2 d\Omega^2,
\end{eqnarray}
where the primes denote internal coordinates.
The coordinate radius of the void is $r_v (t)$ in the external
coordinates and   
${x'}_v(t')$ in the internal coordinates.
They must satisfy 
\begin{eqnarray}
  {x'}_v(t)= a(t) r_v (t).
\end{eqnarray}
Maeda and Sato\cite{rf:MAEDA} and Bertschinger\cite{rf:BERTSCHINGER}
showed that the shell radius expands asymptotically as
\begin{equation}\label{R-evolve}
r_v(t) \propto t^{\beta}~~~{\rm with}~~\beta\sim 0.13 .
\end{equation}
In the following, the subscripts 1 and 2, respectively,  denote
quantities at the time 
the photon enters the void and at the time it leaves.
Primes denote quantities measured by an observer at rest in the
internal coordinate system, and unprimed quantities are measured
by an observer comoving in the universe just outside the shell of the
void. 

\begin{figure}[tbh]
 \epsfxsize = 7cm
 \centerline{\epsfbox{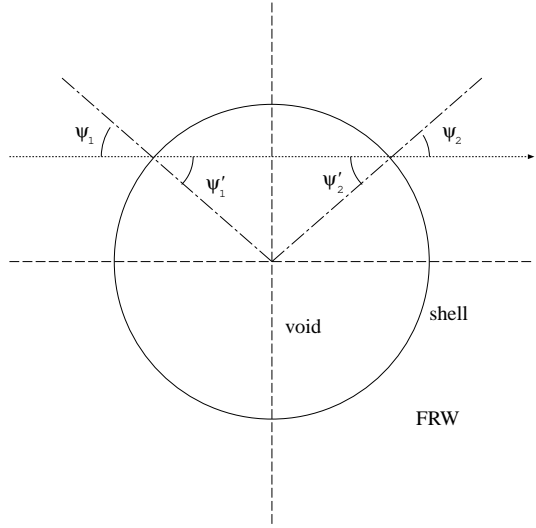}}
\caption{
Cross section of a void and the photon path
through the $\phi = \pi/2$ plane.
The expansion of the void is not shown explicitly.
}\label{fig:angle}
\end{figure}

\begin{figure}[tbh]
 \epsfxsize = 13cm
 \centerline{\epsfbox{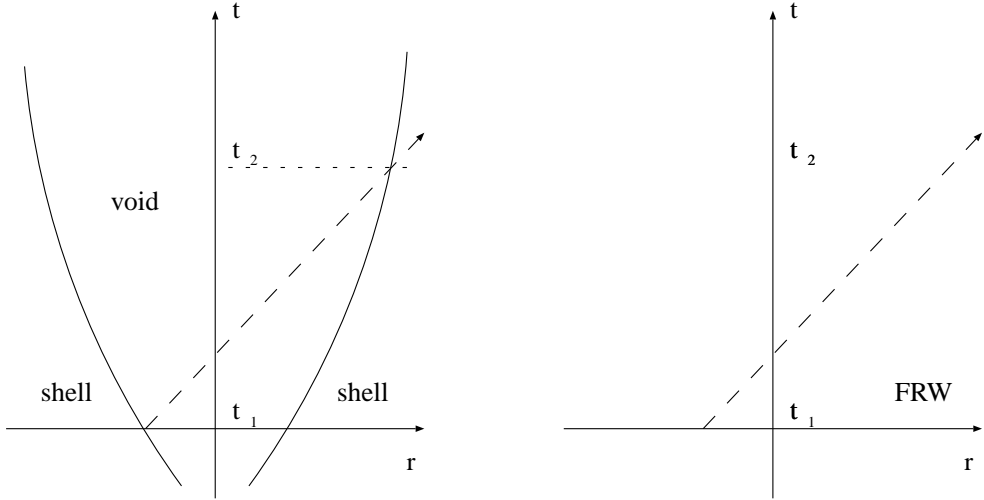}}
\caption{
Trajectory of a dust shell and the path of a light ray
in the void-shell system ({\it left}).
For comparison, we consider a light ray
which travels in the FRW universe for the same time interval
measured outside the void ({\it right}). 
}\label{fig:compare}
\end{figure}

The light ray bundle enters the void at $t=t_1$ 
(corresponding to $t'={t_1}'$ in the internal coordinates). 
Then the light arrives at the other side of the void at $t=t_2$
($t'={t_2}'$).
The arrival time is obtained by solving the equations
\begin{eqnarray}
  t_2'- t_1' &=& x_1' \cos {\Psi_1}' + x_2' \cos{\Psi_2}',\\
  x_1' \sin {\Psi_1}' &=& x_2' \sin {\Psi_2}'.
\end{eqnarray}
(See Fig. \ref{fig:angle} for the definition of the angular variables.)
The redshift and the distance are obtained by 
solving the geodesic equation and the Raychaudhuri equations,
as we do in the following.
We also calculate the distance and redshift
of a null ray 
in the FRW model which has traveled from $t_1$ to $t_2$.
We then compare the distances and redshifts 
for these two models (see Fig.\ref{fig:compare}).
%

Just before entering the shell,
the light ray bundle has 
$k_1$, $d_{\rm A}^- = \dAf$, $\theta_1$ and $\sigma_{AB1}=0$.
At the time it crosses the shell,
we impose Eq. (\ref{eq:theta-jump3}) for $\theta$,
Eq. (\ref{eq:sigma-jump3}) for $\sigma$,
and the following conditions for $k$ and $\dA$:
\begin{eqnarray}
k^a u_a = {k^a}' {u_a}',~~~
k^a n_a = {k^a}' {n_a}',~~~
\dA^+ = \dA^-.
\end{eqnarray}
Note that these conditions for $k^a$ give the same result
as that obtained by the Lorentz transformation adopted by Thompson and
Vishniac.\cite{rf:THOMPSON}\  From 
Eq. (\ref{dA-Min}), the distance 
at the time the ray bundle leaves the void is given by
\begin{eqnarray}\label{eq:dAMin}
  \dA (t_2') = \dAf \left\{ 
    1+{\theta_1'\over 2 k_1'} (t_2' - t_1')\right\}
\times \{ 1 + \Delta\},
\end{eqnarray}
where $\Delta$ is the correction due to the shear term.
This is to be compared with 
the distance in the FRW model given by Eq. (\ref{dA-FRW}).
In the following analysis, 
we assume that the void is much smaller than the Hubble radius,
and we define the small parameter $\eta$ by
\begin{eqnarray}
  \eta \equiv a(t) r_v (t)/t ,
\end{eqnarray}
i.e., the ratio of the radius of the void to the Hubble radius.
We expand all the quantities in powers of $\eta$.

Now let us consider the jumps of $\theta$ and $\sigma$.
The jump of the expansion given by Eq. (\ref{eq:theta-jump3})
is calculated as
\begin{eqnarray}
  \theta_1' = \theta_1 - {4\over 9} f k_1 \gamma_1 
{(1+v_1 \cos  \Psi_1 )^2 \over v_1+\cos\Psi_1}{\eta_1\over t_1},
\end{eqnarray}
where we have defined the velocity and the $\gamma$-factor as
\begin{eqnarray}
v_1 \equiv a(t_1) {dr_1\over dt}= \beta \eta_1,~~~~~
\gamma_1 \equiv(1-v_1^2)^{-\half},
\end{eqnarray}
and the ratio of the shell mass 
to the ``extracted'' mass as
\begin{eqnarray}
f \equiv {{4\pi D a^2 r_v^2}\over (4\pi/3) a^3 r_v^3 \rho_{\rm FRW}}~~~ 
{\rm with}~~\rho_{\rm FRW} = {3H^2\over 8\pi G} ,
\end{eqnarray}
which will be set to unity, but at this stage we keep it 
as a free parameter for later discussion.
We also used
\begin{eqnarray}
u_a k^a = -k_1 \gamma_1 ( 1+ v_1 \cos \Psi_1 ),~~~
n_c k^c = k_1 \gamma_1 (v_1+\cos \Psi_1).
\end{eqnarray}
The shear induced by the shell is calculated using 
Eq.(\ref{eq:sigma-jump3}) as 
\begin{eqnarray}
\sigma_{11}'= - \sigma_{22}'
 = -{2\over 9} f {k_1\gamma_1(1-v_1^2)\sin^2\Psi_1\over v_1 +\cos\Psi_1}
 {\eta_1 \over t_1},
\end{eqnarray}
where we have taken $e_{2a}'= (0,0,0,x_1')$.
Since the increase in the affine parameter 
during the travel through the void is given by
\begin{eqnarray}
 \la - \la_0 = {t_2' - t_1'\over k_1'}
            = 2 {t_1 \over k_1}\cos\Psi_1 \eta_1 + {\cal O}(\eta_1^2),
\end{eqnarray}
the correction term $\Delta$ in Eq. (\ref{eq:dAMin}) is 
\begin{eqnarray}
|\Delta| \sim \sigma^{{AB}}_1 \sigma_{{AB}1} (\la -\la _0)^2 
 = {32\over 81} f^2
{(1-v_1^2)\sin^4\Psi_1\cos^2\Psi_1\over (v_1+\cos\Psi_1)^2}
\eta_1^4 .
\end{eqnarray}
Thus the correction due to the shear is of order $\eta_1^4$.

Now we study the modification of the redshift
due to the existence of the void.
The redshift is defined in the void model by 
\begin{eqnarray}
1+z_{\rm V}\equiv {k_1\over k_2}
\end{eqnarray}
and in the FRW model by
\begin{eqnarray}
1+z_{\rm F} \equiv \left( {t_2\over t_1}\right)^{2\over 3}.
\end{eqnarray}
By a straightforward calculation, we find
\begin{eqnarray}
{1+z_{\rm V}\over 1+z_{\rm F} } -1 = 
\left(
-{16\over 81} \cos^3 \Psi_1 + {9\over 8}\beta \cos \Psi_1
\right) \eta_1^3,
\end{eqnarray}
which agrees with the result of Thompson and Vishniac.\cite{rf:THOMPSON}
\ 
Thus, the leading order of the modification of redshift is 
order $\eta_1^3$. 

Next we study the modification of the angular-diameter distance.
Expanding Eq. (\ref{eq:dAMin}), we find
\begin{eqnarray}
  d_{\rm AV}(t_2) &=& \dAf + \dAf {t_1\over k_1}\theta_1 \cos \Psi_1' \eta_1
  - \dAf {4f\over 9} \eta_1^2 \nonumber\\
&{}& + \dAf {t_1\over k_1}\theta_1 
\left( \beta+{2\over 3} + {2\over 3}\cos^2 \Psi_1' \right)
\eta_1^2 + {\cal O} (\eta_1^3) .
\end{eqnarray}
For the FRW model, we find, from Eq. (\ref{dA-FRW}), 
\begin{eqnarray}
  d_{\rm AF}(t_2) 
&=& \dAf + \dAf {t_1\over k_1}\theta_1 \cos \Psi_1' \eta_1
   - {4\over 3} \cos^2\Psi_1' \eta_1^2 \nonumber\\
&{}& + \dAf {t_1\over k_1}\theta_1 
\left( \beta+{2\over 3} + {2\over 3}\cos^2 \Psi_1'\right)
\eta_1^2 + {\cal O}(\eta_1^3) .
\end{eqnarray}
Thus the difference reads
\begin{eqnarray} \label{eq:difference}
  d_{\rm AV}(t_2) - d_{\rm AF}(t_2) =
 \dAf \left \{ {4\over 3} \cos^2 \Psi_1'- {4f\over 9} \right \}
\eta_1^2  + {\cal O}(\eta_1^3) .
\end{eqnarray}
%
\begin{figure}[tbh]
 \epsfxsize = 11.5cm
 \centerline{\epsfbox{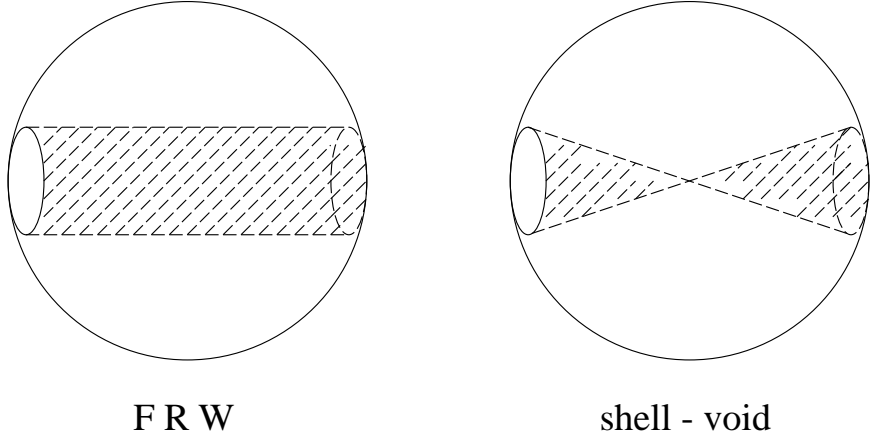}}
\caption{
Schematic picture for the volume of the matter 
which causes the Ricci focusing on the passing photon beam.
In the FRW model ({\it left}), the beam is influenced by 
the matter contained inside the tube. 
In the void-shell sysytem ({\it right}), 
the shell contains only the matter inside the cones. 
}\label{fig:cone}
\end{figure}
Let us focus on the leading term in the difference.
When $\cos \Psi_1'$ is close to unity, i.e. when
the beam passes near the center,
the beam is de-magnified compared with the beam in the FRW model,  
to give a longer distance (we assumed $f=1$).
Contrastingly, when $\cos \Psi_1'$ is close to zero, i.e.
when the impact parameter is large, the beam is magnified.
One can also note that the leading term of $(d_{\rm AV}-d_{\rm AF})$ 
for $\cos \Psi_1' = 1$ vanishes if $f=3$.
Recalling that $f$ denotes the mass fraction of the dust shell 
to the mass that would have been in the void, 
we may say that 
in the void-shell system
the Ricci focusing effect is weaker
than in the FRW model by a factor of ${1\over 3}$. 
This can be explained in the following way
(see Fig. \ref{fig:cone}).
In the FRW universe, the beam suffers Ricci focusing 
by the matter within the tube shaded in the figure.
On the other hand, in the void-shell system,
the area of the shell which causes the Ricci focusing
contains only the matter within the cones
whose tops are at the center of the void.
This explains the factor of ${1\over 3}$.

Our result disagrees with the result obtained by Sato,\cite{rf:SATO}\ 
who argued that the difference is of order $\eta_1$.
The disagreement comes from the junction condition for the distance
at the time that the bundle crosses the shell. Sato assumed that the
derivative of the distance 
with respect to the proper time measured by an observer at the boundary
is continuous across the shell.
However, we cannot read off any clear physical meaning 
that this condition should represent.
Our condition, on the other hand, is obtained 
by integrating the Raychaudhuri equation, whose physical meaning is 
clear.
As shown in the Appendix, the same result is obtained 
by a different method.
Thus we conclude that
the leading term in the difference is of order $\eta_1^2$.
In other words, when a light ray passes only one void 
during its travel from the source to the observer, 
the modification of the distance compared with the whole distance
is of order $\eta_1^2$,
which may be considered to be very small.
However, this means that 
the {\it increase} in the distance during the time that tye crosses
one void is modified by the order of $\eta_1$, since the radius of the
void itself is of the order of $\eta_1$. Moreover, the effect is
cumulative when the beam passes through many voids successively.
Thus, it is interesting to study a multi-void system.
This is the subject of
the next section.

\subsection{Light propagation in multi-void system}
\begin{figure}[hb]
 \epsfxsize = 13cm
 \centerline{\epsfbox{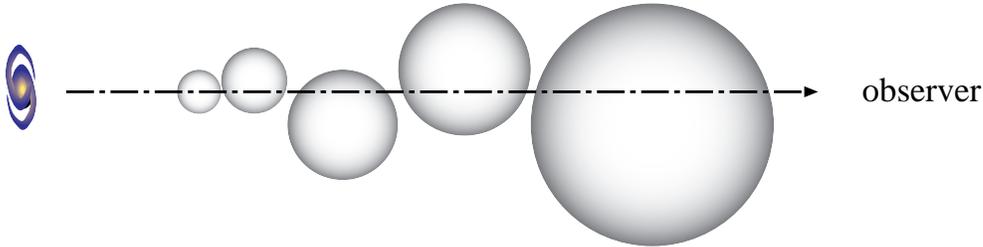}}
\caption{
Photon beam traveling through the universe filled with voids.
We assume that the beam enters another void as soon as it 
leaves one void. 
}\label{fig:voids}
\end{figure}

In this section, 
we study the distance-redshift relation
in a universe filled with voids.
The simple model used here consists of randomly distributed
voids, whose sizes evolve according to Eq. (\ref{R-evolve}).
\footnote{Our approach is similar to the method used by Holz and Wald.
\cite{rf:WALD}\ We have just substituted a void-shell system 
for their particle-hole system.}
We start from a point source at $t=t_1$ in the flat FRW universe.
The light propagates in the FRW universe for 
a time interval $\sim 0.05\times t_1$ 
and then enters the first void surrounded by a dust shell. 
When the ray bundle crosses the shell, 
we calculate the expansion and the null vector 
using the junction condition described in the previous sections.
We perform the same calculation for the time that
 the light leaves the void.
For simplicity, we assume that the light ray enters another shell as
soon as it leaves one void (see Fig. \ref{fig:voids}).
The entering angle (or impact parameter) is treated as a random
variable. At each exiting moment, we can calculate the redshift
$z_{\rm V}$ and  
the angular-diameter distance $d_{\rm AV}$ from the source. 
We continue until $z_{\rm V}$ exceeds a desired value.
We also calculate the angular-diameter distance $d_{\rm AF}$
for a source at the same redshift in the FRW universe. 
By repeating this sequence of calculations,
we accumulate a large number of data (typically $\sim 10^4$)
for $d_{\rm AV}/d_{\rm AF}$. 
We omitted cases when the expansion $\theta$ becomes negative.
The fraction is well below $1\%$ even for the cases where the void
size is about $0.1H_0^{-1}$, and 
thus it has no influence on the following discussion.
We also neglect the shear effect in the subsequent calculations,
since the effect is negligible, as we saw in the previous section.
%

\begin{figure}[tbh]
 \epsfxsize = 11.5cm
 \centerline{\epsfbox{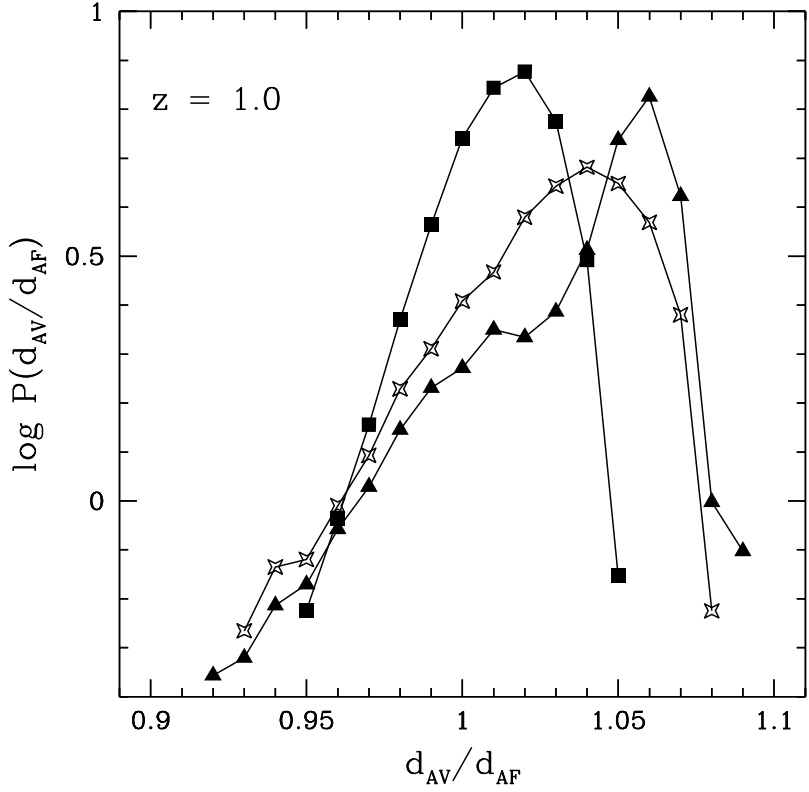}}
\caption{
Probability distribution $P$ of $d_{\rm AV}/d_{\rm AF}$
for samples at redshifts between $1.0$ and $1.1$ .
The normalization of $P$ is such that the fraction of the beams 
which has $d_{\rm AV}/d_{\rm AF}$ in the range $[x, x+dx]$ 
is given by $P(x)dx$.
The radius of the void at present  is
$0.01 H_0^{-1}$(square), $0.05 H_0^{-1}$(cross), 
and $0.1 H_0^{-1}$(triangle) in each model.
}\label{fig:result1}
\end{figure}

\begin{figure}[tbh]
 \epsfxsize = 11.5cm
 \centerline{\epsfbox{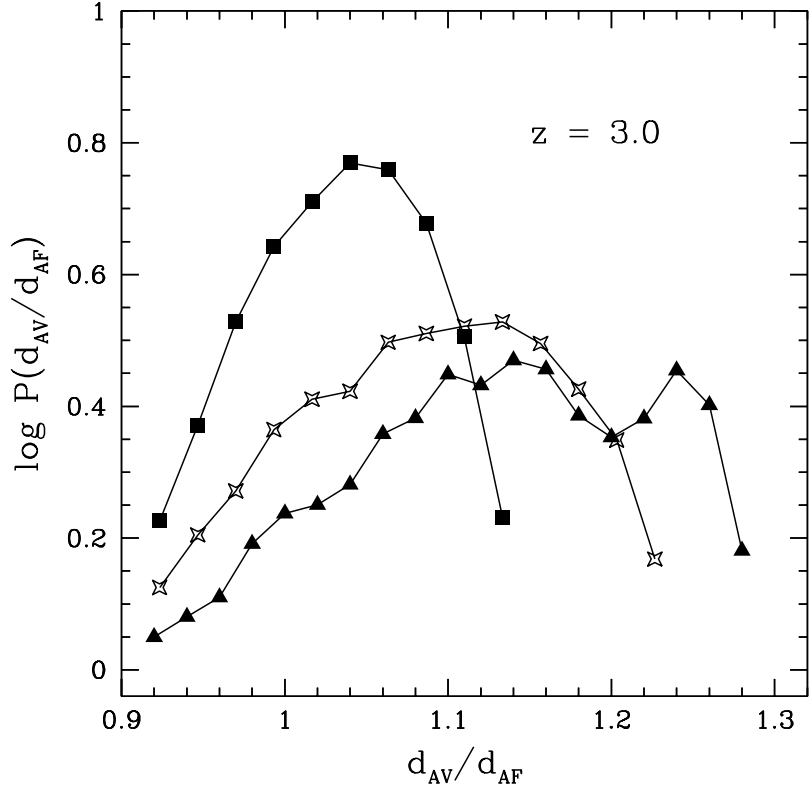}}
\caption{
Probability distribution of $d_{\rm AV}/d_{\rm AF}$
for samples at redshifts between $3.0$ and $3.2$ .
The normalization of $P$ is the same with the previous figure.
The size of the void is labeled by the same symbols as the previous figure.
}\label{fig:result2}
\end{figure}

Figures. \ref{fig:result1} and \ref{fig:result2} display the probability
distribution for $d_{\rm AV}/d_{\rm AF}$.
In Fig. \ref{fig:result1} we plot the distribution for samples at redshift
 between $1.0$ and $1.1$, and in Fig. \ref{fig:result2} for samples
 at redshifts  
between $3.0$ and $3.2$. 
The sizes of the void are adjusted so that the void has evolved 
at the observed time to radii of (approximately) $0.01 H_0^{-1}$,
$0.05H_0^{-1}$ and $0.1H_0^{-1}$,  
where $H_0^{-1}$ is the Horizon radius at the observed time.
We can see that the distribution is peaked 
above unity by $2\%$ $(4\%)$ when the void size is $0.01H^{-1}$ 
and by $4\%$ $(10\%)$ when it is $0.05H^{-1}$ 
for $z\sim 1.0$ ($z\sim 3.0$), although the peak 
is rather broad for a large void size.
The dispersion becomes even larger when the sample redshift
is increased.
Also note that the shape of the distribution is not symmetric.
%
%

When the void radius is $0.1H^{-1}$, 
the distribution seems to have a wavy feature.
In order to see how this occurs, let us study higher order terms 
that we neglected in the previous section.
Expanding Eq. (\ref{eq:dAMin}) up to ${\cal O}(\eta_1^3)$, we find
\begin{eqnarray}
{\cal O}(\eta_1^3)~~{\rm of}~~&&{d_{\rm AV}(t_2')\over d_0} =
 - {8\over 9} f \left(\beta+{2\over 3}\right)\cos\Psi_1 \eta_1^3 \nonumber \\
&& + {t_1\over k_1}{d_0'\over d_0}
         \left[{8\over 3} \cos^3\Psi_1 + 
         \left(4\beta^2 + {14\over 3}\beta-{8\over
           9}\right)\cos\Psi_1\right]\eta_1^3 .
\end{eqnarray}
For the FRW model, we find 
\begin{eqnarray}
{\cal O}(\eta_1^3)~~{\rm of}~~&&{d_{\rm AF}(t_2')\over d_0} =
 - {8\over 3}\left(\beta+{2\over 9}\cos^2\Psi_1\right)\cos\Psi_1\eta_1^3 \nonumber \\
&& + {t_1\over k_1}{d_0'\over d_0}
         \left[{40\over 27} \cos^3\Psi_1 + 
         \left(4\beta^2 + {14\over 3}\beta-{8\over
           9}\right)\cos\Psi_1\right]\eta_1^3 .
\end{eqnarray}
Thus the difference reads
\begin{eqnarray}
{d_{\rm AV}(t_2')- d_{\rm AF}(t_2')\over d_0} &&=
\left({16\over 27}+{t_1\over k_1}{d_0'\over d_0}{32\over 27}
\right)\cos^3\Psi_1 \eta_1^3 \nonumber \\
&&~~~~~~~~~~~~~ - \left[{8\over 9} (f -3\beta) +{16\over 27}f
	\right]\cos\Psi_1 \eta_1^3 .
\end{eqnarray}
This factor takes a small negative value
when $\cos \Psi_1$ is small, i.e. when the beams passes close to the 
void wall.
It decreases (the absolute value increases) as the impact decreases,
then at some point it starts to increase, 
and ends up with a slightly positive value when $\cos \Psi_1 = 1$.
This causes the wavy distributions, which can be seen
more clearly for a larger size of the voids. 

We find the peak of the distribution becomes close to unity
when the void size is small. 
We can crudely interpretate this behavior as follows.
While a lgiht ray bundle passes one void, 
the distance is affected by the amount of $\sim {\cal O}(\eta^2)$.
The number of the voids which the lirhgt ray passes through 
before it reaches the observer is $\dA / r_{\rm v} \sim (1+z)/\eta$. 
Thus, the accumulation of the effects amounts to 
\begin{equation}
\Delta \dA/d_{\rm AF}(z) \sim (1+z)\eta .
\end{equation}
This consideration seems reasonable judging from the figures.

Let us compare our model with the Swiss Cheese (SC) model.
Light propagation in the SC model is studied using a similar approach
by Holz and Wald. \cite{rf:WALD}
According to their work,
when the matter is condensated to point masses
or compact objects, a large fraction of the beams from distant sources
travel without being affected by the matter for most of their travel time.
That is, a large fraction of the beams avoid the Ricci focusing effect,
while a quite small fraction of the beams suffers strong shear focusing.
Thus, the obtained distance for a fixed redshift 
favors a lower density model, even if the background model is a
flat FRW model.
On the other hand, in the void-shell system,
a beam passes through a substantial fraction of the matter,
resulting in a small difference from the FRW distance.
In addition, when the beam successively passes through the voids,
it is affected by almost the same amount of matter in the FRW
universe as we have seen above.
Thus, if the underdensity of the void is compensated
by a surrounding (spherical) shell, 
the difference in distance from the FRW model is small.
Also note that the distributions of distance 
are quite different for the SC model and the void model,
reflecting the difference in the matter distribution.
In principle, we can obtain information concerning the 
inhomogeneities from the scatter or distribution of the
distance-redshift relation, 
though the current accuracy of observation is not sufficient to 
allow this.

We finally note the implications of our results
on the determination of the cosmological parameters
through the distance-redshift relation.
As we have seen, the void structure increases
the distance of a source 
compared with the distance in the FRW model;
in this sense, the cosmological constant or a low density model is 
favored.
However, the magnitude of the modification is
about $2\%$ for a void whose size is $1\%$ of the horizon radius,
which is well below the current error in the estimate
of the absolute magnitude (or distance)
of the standard candles, such as a type Ia super nova.
The scatter around the peak is also well below the current error 
as long as the void is not too large ($ < 1\%$ of the horizon radius).

\section{Summary}
We have studied light propagation properties 
in a spherical void (empty region) embedded in a spatially flat FRW 
universe.
The void is surrounded by an overdense wall,
which we treated as a dust shell.
We considered the null geodesic equation and the Raychaudhuri equation,
and derived the junction condition for the expansion and the shear
across the shell.
Using these, the redshift and the distance were calculated
in the void-shell system,
and we compared the results with the FRW relations.
We have found that the modification of redshift by a single void is 
of order $\eta^3$, 
and the modification of the distance is of order $\eta^2$,
where $\eta$ is the ratio of the void size to the Hubble radius.

We have also discussed the cosmological implications
of voids by considering a universe filled with voids.
The void-filled universe slightly increases the mean distance
and produces a dispersion in the distribution of the distance.
This may work to lower the density parameter
or favor the cosmological constant 
when the cosmological parameters are estimated 
through the distance-redshift relation of distant sources.
However, for samples with redshifts $\sim 1.0$, 
the magnitude of the effect is well below the  
current error in estimating the distance of the standard candles
as long as the voids are within the 
observationally favored size. 

\appendix
\section{Simple Derivation of $\dA$ for a Single Void}

Our analytic method based on the geometric optics equations 
(\ref{eq:expansion-eq}) and (\ref{eq:shear-eq}) is not only useful for 
the present analysis but also can be extended to 
general spacetimes including singular hypersurfaces. For the
model of a void in the flat FRW background, however, we have a simpler 
derivation of the angular-diameter distance which we now present.

\begin{figure}[tbh]
 \epsfxsize = 11cm
 \centerline{\epsfbox{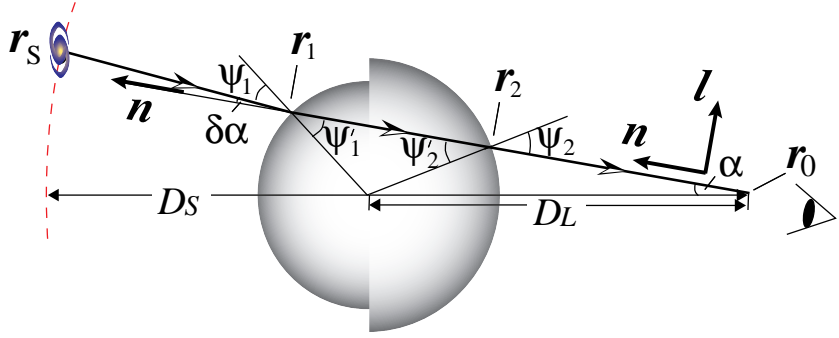}}
 \caption{Cross section of a void on the $\varphi=\pi/2$ plane of the model. We depict the trajectory
of a photon from a light source to an observer. 
Define $\alpha$ as the angle formed between the
direction of observation and the direction of the void's center, 
$\delta\alpha$ as the scattering angle of a
photon, $D_L$ as the comoving distance of the void's center, 
and $D_S$ as the comoving distance of
the light source. We denote each position by a vector symbol $\br$, 
and introduce a vector basis,
$\{\bn(\alpha),\bl (\alpha)\}$, in the two-dimensional comoving space.}
 \label{fig:A1}
\end{figure}

Thompson and Vishniac derived not only the redshift deviation (3.19)
but also the scattering angle of a photon: \cite{rf:THOMPSON}
\beq
\delta\alpha\equiv -\Psi_1+\Psi_1'+\Psi_2'-\Psi_2 
=\frac49\eta_2^2\sin(2\Psi_2).
\eeq
As depicted in Fig. \ref{fig:A1}, we define $\alpha$ as the angle
formed between the  
direction of observation and the direction of the void's center, 
$D_L$ as the comoving distance of the void's center, and $D_S$ as the comoving 
distance of the light source. Hereafter the subscript $S$ denotes a
quantity at a  
light source. In the two-dimensional comoving space we denote 
each position by the vector symbol $\br$ and introduce a vector 
basis, $[\bn(\alpha),\bl(\alpha)]$. The trajectory of a 
photon between a source and an observer was obtained in 
Ref. 14). It is given by
\bea
\br_S-\br_0&=&3t_0
\left[(1-\sqrt{a_S})\bn+\left(1-\sqrt{a_S}-{D_L\over3t_0}\right)\delta\alpha\bl\right] \nonumber\\
&\equiv&D_S\bn+D_{LS}\delta\alpha\bl,
\eea
where $D_{LS}\equiv D_S-D_L$. In the absence of a void, this relation 
reduces to
\beq
(\br_S-\br_0)_b=3t_0(1-\sqrt{a_S})\bn\equiv D_S\bn,
\eeq
where the subscript $b$ denotes a background unperturbed quantity.
\begin{figure}[tbh]
 \epsfxsize = 11cm
 \centerline{\epsfbox{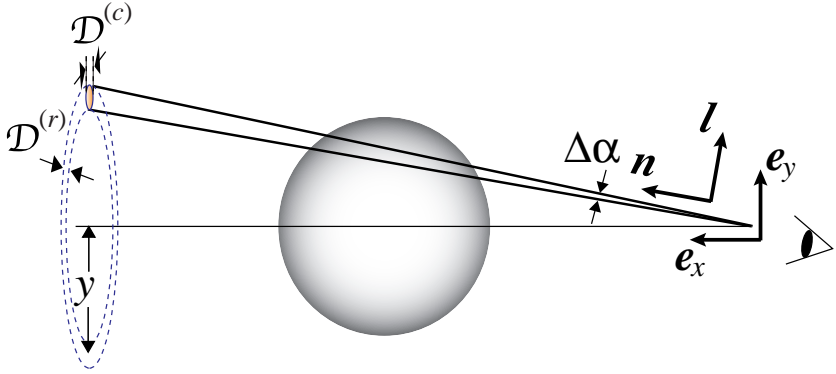}}
 \caption{Schematic picture of a null geodesic congruence. $ \cd^{(r)}$ 
 and $\cd^{(c)}$ denote, respectively, the radial component and the circumferential 
component of the proper distance at a source for a fixed angular diameter.}
 \label{fig:A2}
\end{figure}

The angular-diameter distance of a light source is originally defined as
\beq
\dA\equiv{\cd\over\delta},
\eeq
where $\cd$ is the proper distance across the source and $\delta$ is 
its angular diameter. Here we calculate the manner in which $\cd$ is
changed by a void for a 
fixed $\delta$, considering the radial direction from the 
void's center and the circumferential direction separately
(see Fig. \ref{fig:A2}).
The radial component $\cd^{(r)}$ and the circumferential 
component $\cd^{(c)}$ are written as
\bea
\cd^{(r)}&=&\left|{d(\br_S-\br_0)\over d\alpha}\right|\Delta\alpha, \\
\cd^{(c)}&\propto&y=(\br_S-\br_0)\cdot\be_y,
\eea
where $y$, $\be_y$ and $\Delta\alpha$ are defined in Fig. \ref{fig:A2}.
A straightforward calculation leads to
\bea
{\Delta \dA^{(r)}\over \dA^{(r)}} = {\cd^{(r)}\over \cd^{(r)}_b}-1
&=&{D_{LS}\over D_{S}}{d(\delta\alpha)\over d\alpha}, \\
{\Delta \dA^{(c)}\over \dA^{(c)}} = {\cd^{(c)}\over \cd^{(c)}_b}-1
&=&{D_{LS}\over D_{S}}\delta\alpha\cot\alpha.
\eea
Thus, the order of magnitude of the modification of the distance
is essentially determined by the scattering 
angle $\delta\alpha$, which is of order $\eta_2^2$.
\acknowledgements
 The authors thank Kouji Nakamura
 of Keio University for valuable discussions.
 N. Sugiura, D. Ida and N. Sakai are supported by
research fellowships of the Japan Society for 
the Promotion of Science for Young Scientists.

\end{document}